\documentclass[10pt,letterpaper]{article}
\usepackage{opex3}
\usepackage{mathtools}
\usepackage{color}

\begin{document}

\title{3D Beam Reconstruction by Fluorescence Imaging}

\author{Neal Radwell, Mordjane A. Boukhet and Sonja Franke-Arnold}

\address{SUPA, School of Physics and Astronomy, University of Glasgow, Glasgow G12 8QQ, UK}

\email{neal.radwell@glasgow.ac.uk} 


\begin{abstract}
We present a technique for mapping the complete 3D spatial intensity profile of a laser beam from its fluorescence in an atomic vapour.  We propagate shaped light through a rubidium vapour cell and record the resonant scattering from the side. From a single measurement we obtain a camera limited resolution of 200 x 200 transverse points and 659 longitudinal points. In constrast to invasive methods in which the camera is placed in the beam path, our method is capable of measuring patterns formed by counterpropagating laser beams.  It has high resolution in all 3 dimensions, is fast and can be completely automated. The technique has applications in areas which require complex beam shapes, such as optical tweezers, atom trapping and pattern formation.
\end{abstract}

\ocis{(000.0000) General.} 


\section{Introduction}

Laser beams with increasingly intricate and complex profiles have become interesting for a range of applications. Particles from the micron size range down to single atoms can be trapped by the dipole forces produced from light beams \cite{grier2003revolution,grimm2000optical}. These forces have been exploited in optical tweezers \cite{bowman2013optical}, allowing micro manipulation of beads and biological matter which would otherwise be impossible. Atoms can also be trapped by the same forces and have the further advantage that their strong resonances allow tuning of the force as a function of the detuning.

With the advent of Spatial Light Modulators (SLMs) it has become possible to generate a wide variety of beam profiles, expanding the possibilities for trapping and guiding. Beam propagation and in particular focussing can result in an additional modification of the beam profile along the propagation axis. Complex 3D beam shapes have been proposed \cite{arlt2000generation,franke-arnold2007,zhang2010generation,arnold2012extending} and implemented in atom trapping \cite{ozeri1999long,xu2010trapping} and optical tweezers \cite{padgett2011tweezers,lee2012optical,bowman2010particle,whyte2005experimental}. In order to verify the accuracy of the beam generation however, traditional methods would require stepping a camera in the beam path followed by reconstruction \cite{leach2005vortex,romero2011entangled}. This method suffers from several drawbacks: The beam axis distance can be difficult to measure accurately, the process is manual and slow and it is clearly ineffective at imaging beam structures created by interference of counter-propagating laser beams. We present here an alternative method which avoids these issues.

\begin{figure*}
\centerline{
\includegraphics[width=\columnwidth]{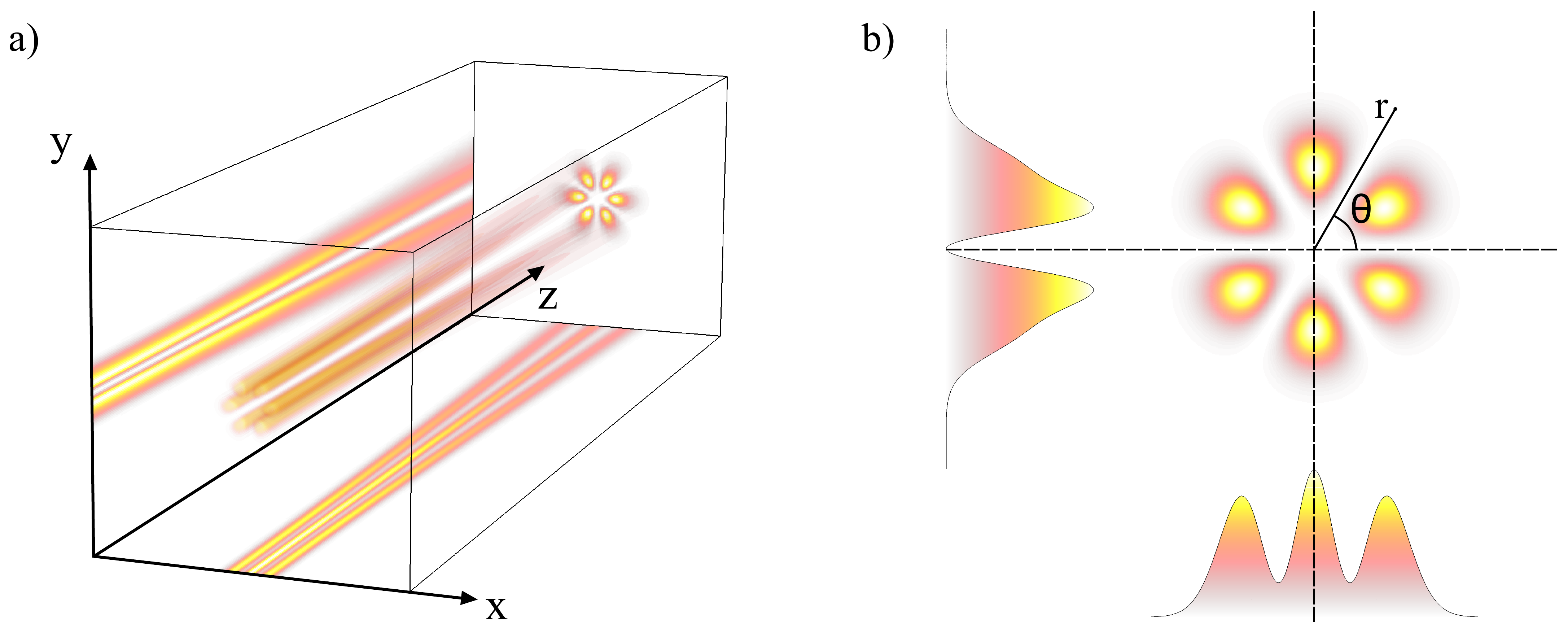}
}
\caption{\label{fig:Axes}a) Illustration of a laser beam including the axis convention used in this paper. The beam shown is a superposition of two Laguerre Gaussian beams with $l$ numbers 3 and -3. The images on the xy, xz and yz planes are the projections of this composite beam onto these planes. b) Cut of the same beam at $z=0$ illustrating the Radon transform: the two profiles are the projections of the 2D pattern in the x and y planes.}

\end{figure*}

\section{Summary of Technique}

Our technique is based on tomographic reconstruction of images taken via fluorescence imaging. If a laser beam is shone through a fluorescent medium the amount of  fluorescence depends on the intensity of the laser beam. In the case of an atomic gas the two-level photon scattering (fluorescence) rate is;

\begin{equation}
R_{\rm sc} =\frac{\Gamma}{2} \frac{(I/I_{\rm sat})}{1+(I/I_{\rm sat})+4(\Delta/\Gamma)^2},
\label{eqn:Rscatt}
\end{equation}

where $\Gamma$ ($2\pi\times6.06$~MHz) is the inverse of the upper state lifetime, $I$ is the laser intensity, $I_{\rm sat}$ (1.6~mw/cm$^2$) is atom specific saturation intensity and $\Delta$ is the laser detuning. 
For laser beams well below saturation, as used in the reported experiments here, the scattering rate becomes directly proportional to the laser intensity. It is therefore possible to measure the intensity of a laser beam by shining a resonant laser through an atomic gas and detecting the emitted fluorescence from the side with a camera. This recorded fluorescence image contains the projection of the light profile onto e.g. the yz plane (see Fig.~\ref{fig:Axes}a) without any information on the x dependence, but the full beam profile can be recovered tomographically.

Tomography is the method to reconstruct 2 dimensional information from 2 or more 1 dimensional projections. A single column of pixels within the camera image provides the projection of the fluorescence along the x-direction (or in cylindrical coordinates at a particular angle $\theta$) at a particular z-position.  
If we were to rotate the camera around the z axis, we would have access to more projections at more angles. From this combined data we can tomographically reconstruct the 2D xy plane at the chosen z-position. By performing this analysis on all columns in each image, and {\it stacking} the retrieved profiles along the z-direction we recover the full 3 dimensional fluorescence distribution.  For practical reasons, instead of rotating the camera around the beam, we instead rotate the beam itself with a fixed camera position.

\section{Tomography}

The operation of projecting 2D data along an axis to produce a 1D profile (See Fig.~\ref{fig:Axes}b) is called a Radon transform \cite{Radon1}, while the operation to return to the 2D data from the 1D profiles is the inverse Radon transform. There are several techniques to compute this inverse, the most common being the back projection method. This method comprises 4 simple steps. First, each 1D projection is filtered to correct for the oversampling of the central pixels. The ideal filter is simply a ramp filter in frequency space, however the filter may be optimised to enhance the important features in an image. The filtered 1D projections are then converted to a 2D image by simply copying the value in each column to every row (i.e. the values are "smeared" from 1D to 2D). Next each 2D image is rotated by the angle at which the projection was taken and finally all of these images are added together.

\begin{figure*}
\includegraphics[width=\columnwidth]{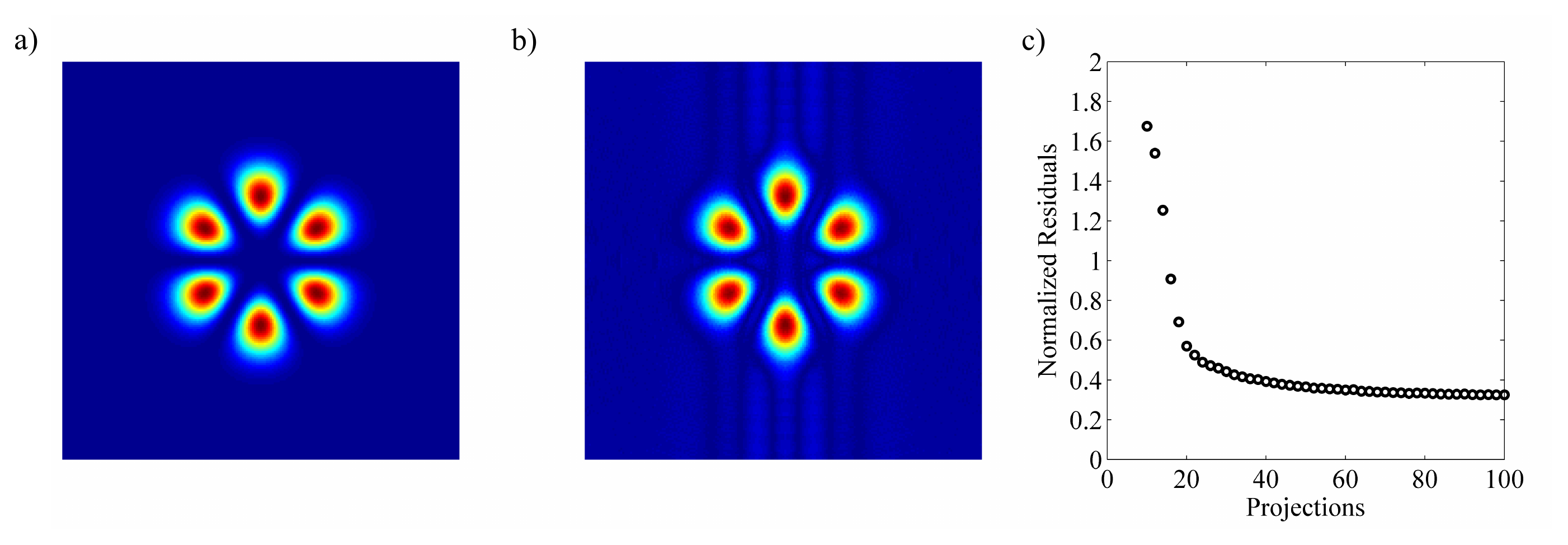}
\caption{\label{fig:Reconstruction_Example}Sample Reconstruction. a) Calculated transverse profile of the beam shown in Fig. 1. b) Reconstruction of the profile in (a) from 30 1D projections, performed in 0.15~s. c) Analysis of the accuracy of the reconstruction with the number of projections. The residuals are the absolute difference between the original profile and the reconstruction, summed over all pixels and divided by the total of all pixels in the original image.}
\end{figure*}

The reconstruction is performed in Matlab and runs extremely quickly. To test the reconstruction we calculate an ideal 2D profile, take 1D projections from this and then reconstruct the 2D profile from the projections. The test profile is shown in Fig.~\ref{fig:Reconstruction_Example}a and a sample reconstruction in Fig.~\ref{fig:Reconstruction_Example}b. The shown reconstruction is formed from 30 projections at 200 x 200 (x,y) resolution, and was generated in 0.15 seconds. The performance of the reconstruction  has been analysed for different numbers of projections, with the results shown in Fig.~\ref{fig:Reconstruction_Example}c. Increasing the number of projections increases the accuracy of the reconstruction at the cost of longer reconstruction times,where the time scales linearly with the number of projections (5ms per projection). We suggest 30 projections as a workable balance between speed and accuracy. For this proof of principle experiment however, we favour accuracy at the expense of speed and therefore the following results have been performed with a large number (116) of projections. 

\begin{figure*}
\centerline{
\includegraphics[width=8cm]{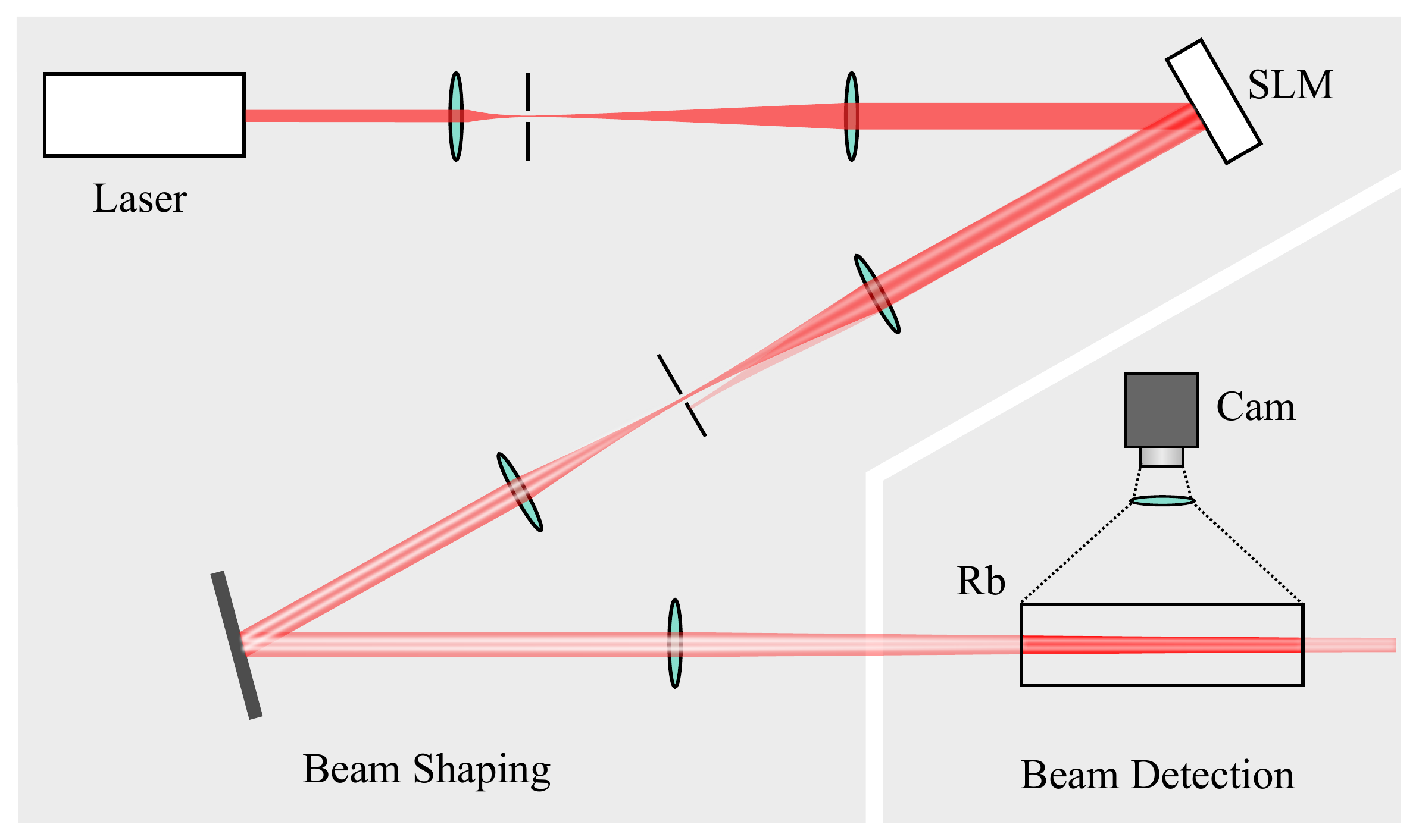}
}
\caption{\label{fig:Setup}Experimental Setup. Beam shaping; The laser mode is cleaned by a pinhole, expanded and then shaped by the SLM. A Fourier filter selects the first order diffracted spot from the SLM hologram. Beam Detection; The fluorescence from the rubidium is imaged and magnified by a single lens and captured on a camera.}
\end{figure*}

\section{Experimental Results}

 To test our beam reconstruction we have set up an experiment as shown in Fig.~\ref{fig:Setup}. The experiment is composed of two main sections, beam creation and beam detection.

The laser source is an external cavity diode laser, tuned to $780$\,nm,  resonant with the Doppler broadened $^{85}$Rb F=2 to F'=3 transition for maximum fluorescence. The mode is cleaned by a pinhole and then expanded. An SLM (Hamamatsu LCOS) is used to generate the shaped beams, allowing us to create arbitrary phase and intensity profiles. We refract the beam off an intensity modulated holographic grating \cite{bowman2011optimisation} and select the first order diffracted beam with an aperture placed in the Fourier plane of the SLM. The beam is then collimated again and sent to the beam detection section.

The beam is directed through a standard rubidium cell in the beam path. The fluorescence emitted from the side of the cell is imaged onto a camera  (Prosilica GC660) by a single lens.

For convenience we rotate the beam by simply turning the pattern on the SLM, however a more general approach would be to use a series of two Dove prisms to rotate the beam, thereby allowing detection of arbitrary beams \cite{leach2002measuring}.

\begin{figure*}
\includegraphics[width=\columnwidth]{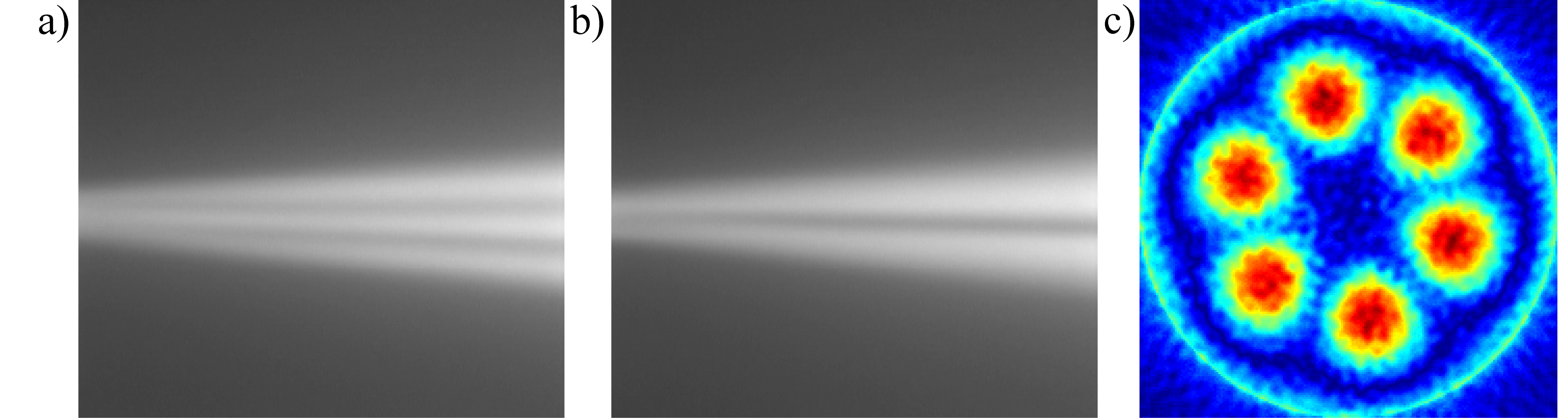}
\caption{\label{fig:Sample_Frames}a) Fluorescence image of the LG superposition of l=3,-3. b) Same as in a) with the beam rotated by an angle of 90$^\circ$. c) Sample reconstruction from 116 projections taken from the same experimental dataset from which a) and b) are taken.}
\end{figure*}

The data shown is of a superposition of two  Laguerre-Gaussian (LG) beams.  A single LG beam is parameterised by the azimuthal and radial numbers $l$ and $p$ respectively. For simplicity we restrict our pattern to have p=0. A single LG beam has an intensity minimum in the centre, the size of which increases proportional to $\sqrt{l/2}$. A superposition of 2 beams, with different l numbers produces a 'petal pattern' with a number of petals equal to the difference in l numbers. We use a superposition of l=3 and l=-3, resulting in the pattern shown in Fig.~\ref{fig:Reconstruction_Example}a, giving 6 'petals'. This pattern is then rotated and a video is recorded by the camera. Sample frames, separated by $90$ degrees, are shown in Fig.~\ref{fig:Sample_Frames}a and b.

\begin{figure*}
\includegraphics[width=\columnwidth]{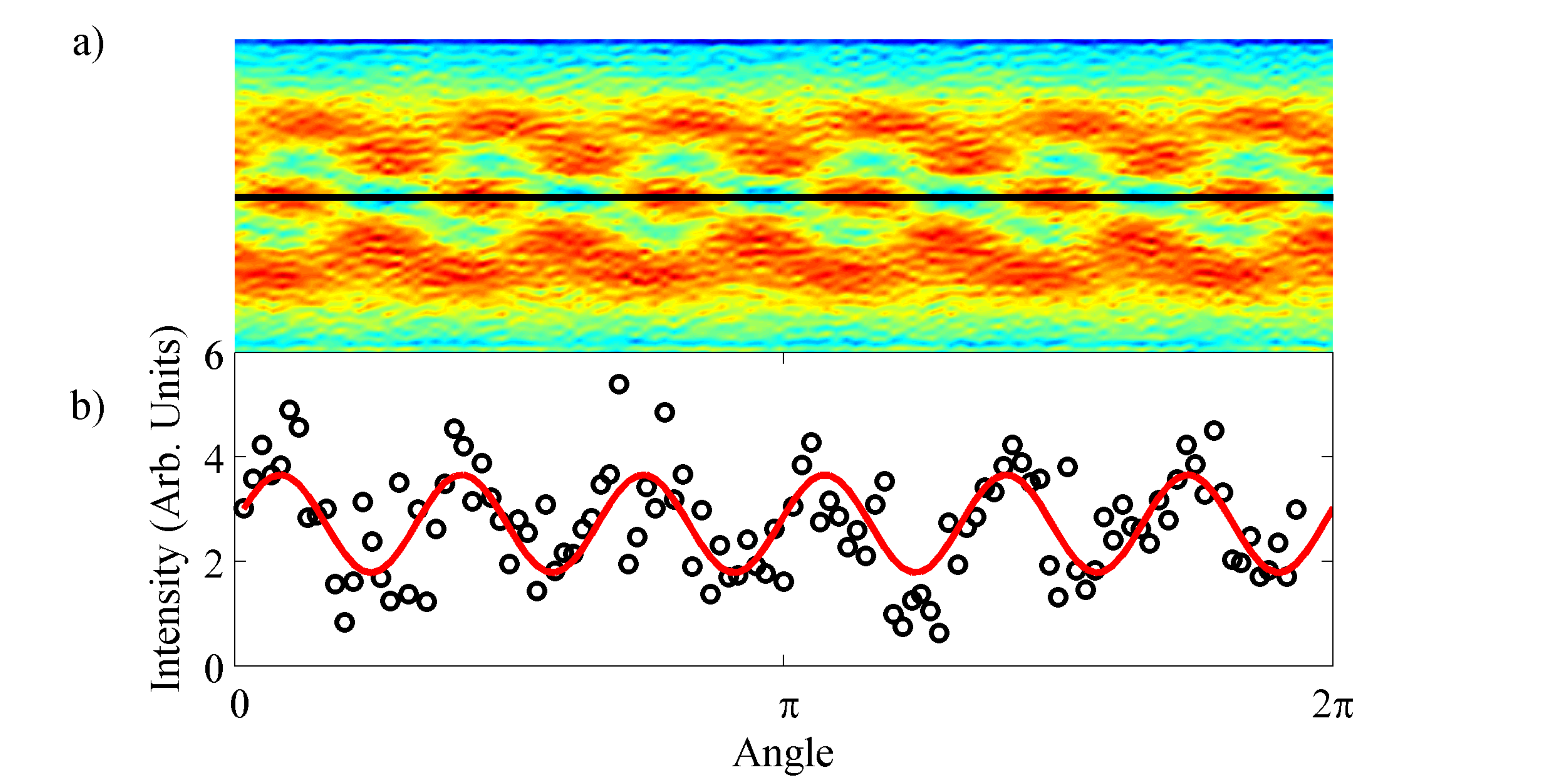}
\caption{\label{fig:Sinogram}a) Sinogram formed by placing all 1D projections as the columns of an image. The data is taken from a single column of each image in the same experiment as in Fig.~\ref{fig:Sample_Frames} b) Circles are a cut through the black line in Fig.~\ref{fig:Sinogram}a, line is a sine fit to the data, from which the relative angle of each projection can be inferred.}
\end{figure*}

To perform the tomographic reconstruction from the experimental data it is necessary to know which angle the beam is rotated in each frame. This can be achieved either by rotating by a known amount from frame to frame, or in our case, by  extracting the angles from the data. A sinogram can be created by stacking all of the 1D projections next to one another, thereby forming an image with $\theta$ along the x axis. An example of which is shown in Fig.~\ref{fig:Sinogram}a. This sinogram has an inherent periodicity which is related to the rotational symmetry of the beam, therefore by fitting a sine function to the x axis of Fig.~\ref{fig:Sinogram}a, we can infer the number of frames between 0 and 2$\pi$.


The result of the reconstruction for a single column of pixels is shown in Fig.~\ref{fig:Sample_Frames}c. Reconstruction for every column results in a 3 dimensional matrix with values corresponding to the fluorescence of the atoms.

\section{Visualisation}

\begin{figure*}
\includegraphics[width=\columnwidth]{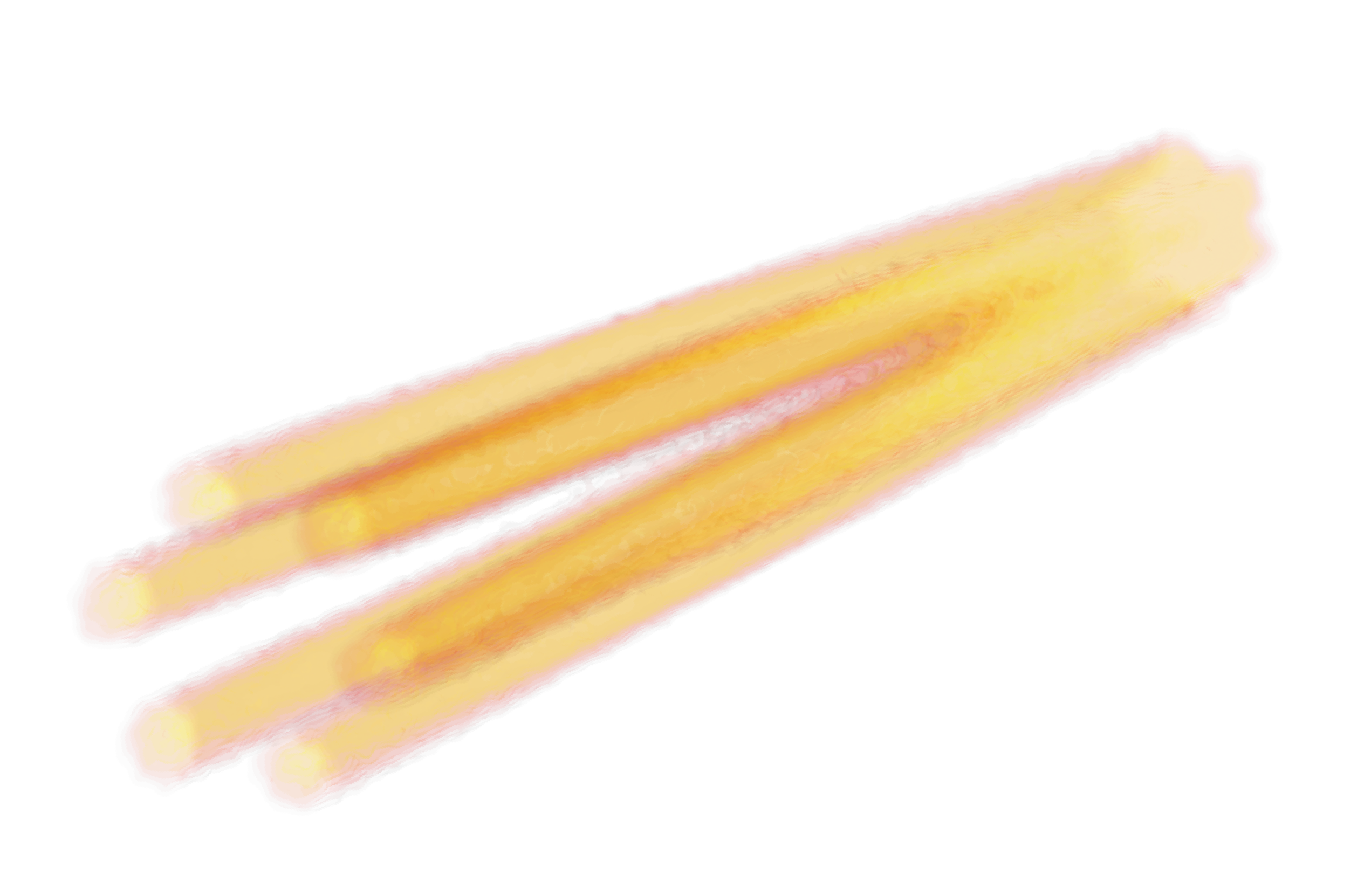}
\caption{\label{fig:3D_Reconstruction} Full 3 dimensional beam reconstruction}
\end{figure*}

The full 3D intensity profile can be visualised with our Matlab program which renders isosurfaces and plots these with a colour and transparency which is related to the intensity of the isosurface. An example plot with ten isosurfaces  between  45~\% and 80~\% of the maximum intensity is shown in Fig.~\ref{fig:3D_Reconstruction}. The visualisation technique has also been expanded to create fly-by videos as well as images and video capable of being displayed in 3D on compatible displays.  We note that in fact the 3D intensity profile can also be observed directly by looking at the scattered light  through an IR viewer.

\section{Discussion}

The resolution of the 3D profiles is generally limted only by the camera.  In order to reduce computation time, here we have used only 200x200x659 pixels of the 494x494x659 camera resolution.  The accuracy of the reconstruction is limited by the number of projections and the noise. The number of projections increases the accuracy of the reconstruction at the expense of longer measurements and calculation time but is already very high for only 30 projections as demonstrated in Fig.~\ref{fig:Reconstruction_Example}c .  
Regarding the noise, a single background measurement can be taken at the start of the experiment, and used to remove background light signals from all further measurements.   This is particularly effective at improving the final signal to noise ratio, due to the high noise sensitivity of the back projection method used to reconstruct the beam.

We next consider high speed measurements or video framerates. Due to the 120~Hz framerate of our camera, a single measurement (at 30 projections) would take 250~ms, however our system is limited by the beam rotation speed which for our SLM is capped at 60~Hz, resulting in 30 projections in 500~ms. A setup involving rotating Dove prisms may be able to outperform the SLM.  
Either method could allow taking 3D measurements at a rate around 1~Hz, so that the limiting factor will still be the reconstruction time. A single z position at 30 projections has a reconstruction time of 150~ms, limiting the z resolution to $<$10 for 1Hz video. Alternatively, z resolution could be increased at the cost of transverse resolution, either by restricting the number of projections or pixels used.

\section{Conclusion}

We have demonstrated a high resolution, simple, fast experimental method which maps the entire 3 dimensional structure of an arbitrary laser beam. As a proof of concept we have generated a superposition of co-propagating LG beams and reconstructed and plotted the results. The method benefits from a far higher resolution in the propagation direction than other invasive methods, without loss of resolution in the transverse dimension. The method is completely automated and  capable of producing a single plane reconstruction in only 150~ms, limited by computation time. 

\end{document}